# HIGHER ORDER LAMB-CHAPLYGIN VORTICES, CHAPLYGIN ANTISYMMETRIC VORTICES, AND ATMOSPHERIC BLOCKING

**Gerald E. Marsh**
**Argonne National Laboratory (Retired)**
**gemarsh@uchicago.edu**

**ABSTRACT**

The equations describing the two-dimensional vortices first described by Chaplygin in 1899 and 1903 are shown to have solutions of higher order that differ from a simple dipole. It is also shown that for these higher-order vortices a very small asymmetry dramatically changes the topology of the stream function. These vortices could be important for understanding the phenomenon of atmospheric blocking.

## Introduction

The Lamb-Chaplygin vortex dipole consists of two regions of opposite vorticity. Lamb[1] described a dipole that is a subset of those described by Chaplygin,[2,3] and the designation Lamb-Chaplygin will often be used here for the symmetric vortex thereby distinguishing it from the Chaplygin antisymmetric vortex. This vortex structure has also been discussed in the book by Batchelor.[4] An extensive history and a concise account of the Chaplygin solutions, as well as an analysis of a non-symmetric vortex dipole moving along a circular trajectory, has been given by Meleshko and van Heijst.[5] Their set of references is quite comprehensive.

Both Lamb and Chaplygin looked for steady motion solutions to the two-dimensional Euler equation for inviscid incompressible flow of a circular vortex of radius $a$ with constant translation velocity $U$. They then superimposed on the fluid a uniform velocity $-U$, thus transforming the problem to that of a stationary one of a vortex cylinder in a fluid with uniform velocity at infinity. In the region outside $r = a$, the flow is assumed to be irrotational; i.e., a potential flow that is the gradient of a scalar function.

Inside the vortex cylinder, the stream function must then satisfy

$$\frac{\partial^2 \psi}{\partial r^2} + \frac{1}{r}\frac{\partial \psi}{\partial r} + \frac{1}{r^2}\frac{\partial^2 \psi}{\partial \theta^2} = F(\psi).$$

(1)

Both Lamb and Chaplygin choose the arbitrary function $F(\psi)$ to have a liner relationship to the vorticity so that

$$F(\psi) = -\omega = -k^2 \psi,$$

(2)

where $\omega$ is the vorticity and $k$ is a constant. The equations to be solved are then

$$\frac{\partial^2 \psi}{\partial r^2} + \frac{1}{r}\frac{\partial \psi}{\partial r} + \frac{1}{r^2}\frac{\partial^2 \psi}{\partial \theta^2} = -k^2 \psi, \qquad 0 \leq r \leq a,$$

$$\frac{\partial^2 \psi}{\partial r^2} + \frac{1}{r}\frac{\partial \psi}{\partial r} + \frac{1}{r^2}\frac{\partial^2 \psi}{\partial \theta^2} = 0, \qquad r > a, (where \ \omega = k = 0).$$

(3)

The solution to Eq. (1), assuming the form for $F(\psi)$ given in Eq. (2), is in general a product of a sum of Bessel and Hankel functions with a sine or cosine function. The fact that the vortex dipole is asymmetrical with respect to the direction of flow implies that the sine function should be chosen. In addition, because the solution must be finite in the domain $r \leq a$, it is the Bessel functions $J_n(kr)$ that are of interest.

Consider the first of Eqs. (3). To solve it one can use the method of separation of variables where $\psi$ is written as $\psi(r, \theta) = R(r)\,\Theta(\theta)$. $R(r)$ is then Bessel's equation for $J_n(kr)$ and $\Theta(\theta)$ has two solutions $A_n \sin(n\theta)$ and $B_n \cos(n\theta)$. The separation constant is $n^2$. The boundary conditions at $r = a$ must then be imposed as discussed in the next section. Without the imposition of those boundary conditions, the solutions of the first of Eqs. (3) can be quite complex as shown in the figures below:

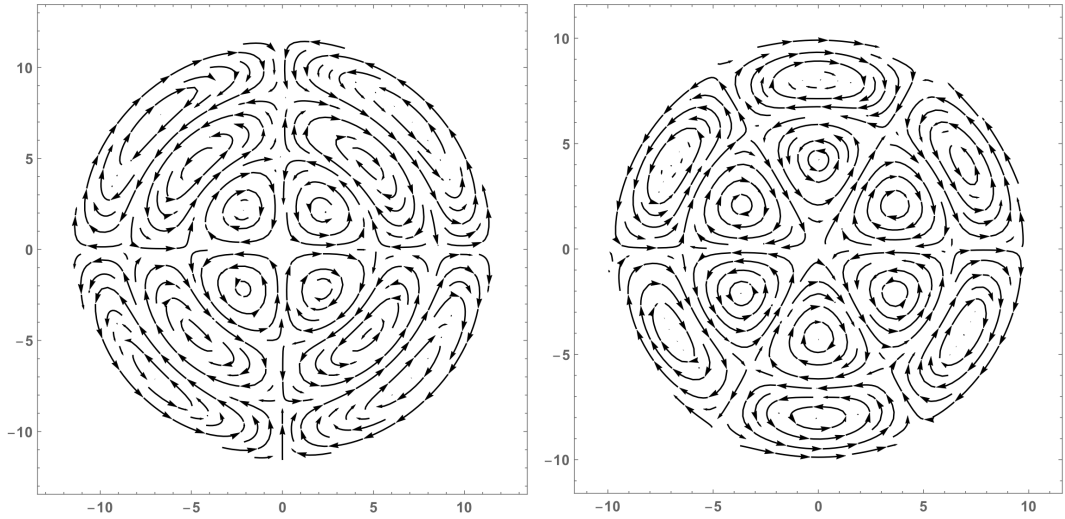

The figure on the left above is for the $n = 2$ Bessel function and the one on the right for $n = 3$. Imposing the boundary conditions below requires the that only $n = 1$ be used. As the streamline plots above make clear, it would not be possible to satisfied the boundary conditions for the flow around a cylindrical boundary as given by Eq. (5) below.

Once the stream function is determined, the velocities can be determined from
$$u_r = \frac{1}{r}\frac{\partial \psi}{\partial \theta}, \quad u_\theta = -\frac{\partial \psi}{\partial r}.$$

(4)

**Form of the stream function in the two domains $r \leq a$ and $r > a$**

The exterior stream function, usually for the flow past a material cylinder of radius $a$, is the superposition of a uniform flow and a doublet and is given by

$$\psi_e = Ur\left(1 - \frac{a^2}{r^2}\right)\sin\theta \qquad r > a.$$

(5)

Following Lamb and Chaplygin, the Bessel function $J_1(kr)$ will be used for the interior solution for $r \leq a$, where the "cylinder" no longer necessarily represents a material cylinder; the higher order functions $J_n(kr)$ for $n > 1$ even if they could be used offer little qualitatively new here (they look like stretched out versions of $J_1$). The interior solution can then be taken to be

$$\psi_i = C J_1(kr)\sin\theta \qquad r < a.$$

(6)

To determine the constant $C$, the boundary conditions at $r = a$ must be used. These are that tangential or azimuthal velocity be equal at $r = a$ and that $\psi_e(a) = \psi_i(a)$. The tangential velocity is $u_\theta = -\frac{\partial \psi}{\partial r}$ so that at the boundary one must have $\frac{\partial \psi_i}{\partial r}\big|_a = \frac{\partial \psi_e}{\partial r}\big|_a$. The condition $\psi_e(a) = \psi_i(a)$ means that $J_1(ka) = 0$; i.e., that $ka$ is a root of the Bessel function $J_1$. Substituting $\psi_i$ and $\psi_e$ into the boundary condition for the tangential velocity gives the constant $C$ as

$$C = \frac{2U}{k J_1{}'(ka)} = \frac{2U}{k J_0(ka)},$$

(7)

where the prime indicates differentiation with respect to $r$.

The full stream function for the Lamb-Chaplygin vortex dipole is then

$$\psi_i = \frac{2U}{k J_0(ka)} J_1(kr)\sin\theta \qquad 0 \leq r \leq a.$$

(8)

To be consistent with the literature, $k$ will be written as $b/a$ with $b$ a root of $J_1(b) = 0$, so that Eq. (8) becomes

$$\psi_i = \frac{2Ua}{b\,J_0(b)} J_1\!\left(\frac{br}{a}\right) \sin\theta \qquad 0 \le r \le a.$$

(9)

Except where noted, however, $a$ will generally be set equal to $b$ since the distinction between the two came from the flow around a physical boundary cylinder, which need not be the case. The appropriate boundary condition when no "boundary cylinder" is present will be introduced below.

**The Lamb-Chaplygin vortex dipole and its higher order generalization**

There will be three ways used to represent vortices: plots of the streamlines as a function of the polar coordinates $r$ and $\theta$,[†] parametric plots where the parameters are the numeric values of the stream function and the radius, and contour plots. The streamline plot is used in Fig. 1.

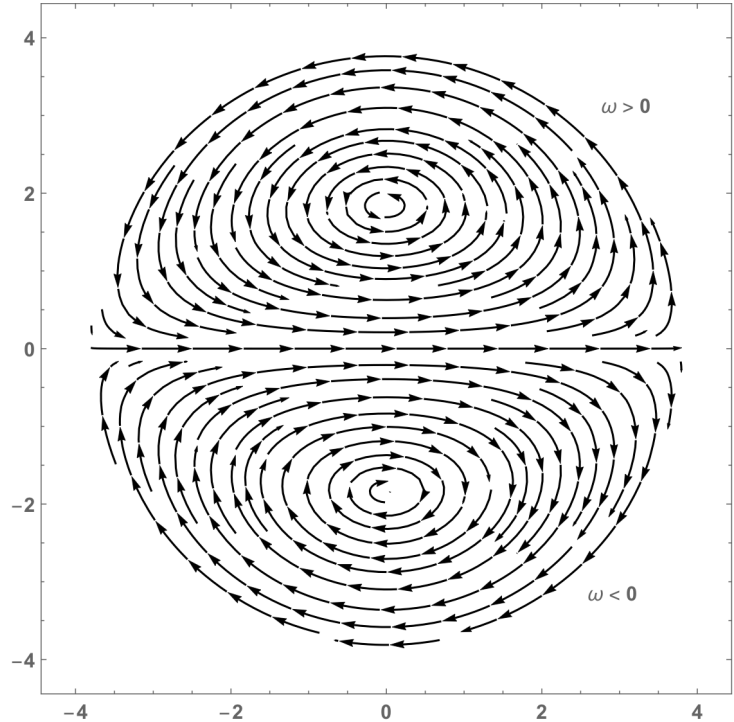

Figure 1. The streamline plot of Lamb-Chaplygin dipole vortex in Cartesian coordinates. Here $U = -0.2$ and $b$ is the smallest positive root of $J_1(b) = 0$, which is 3.8317. The positive vorticity corresponds to the upper portion of the plot and the negative vorticity to the lower.

---

[†] For such streamline plots, since the instantaneous velocity is always tangent to the streamline, the velocity may be represented by the density of streamlines. For this reason, streamlines will be close together where the velocity is large and farther apart in regions where the velocity is slower.

The velocities in the stream plot of Fig. 1, where $a$ has been set equal to $b$, are given by Eq. (4) and are

$$u_r = \frac{2Ua}{br\, J_0(b)} \cos\theta\, J_1\left(\frac{br}{a}\right),$$

$$u_\theta = -U\, \frac{J_0\left(\frac{br}{a}\right) - J_2\left(\frac{br}{a}\right)}{J_0(b)} \sin\theta.$$

(10)

These velocities are used in Figure 2 to make a streamline plot of both $\psi_i$ and $\psi_e$ for a few values of $\psi$.

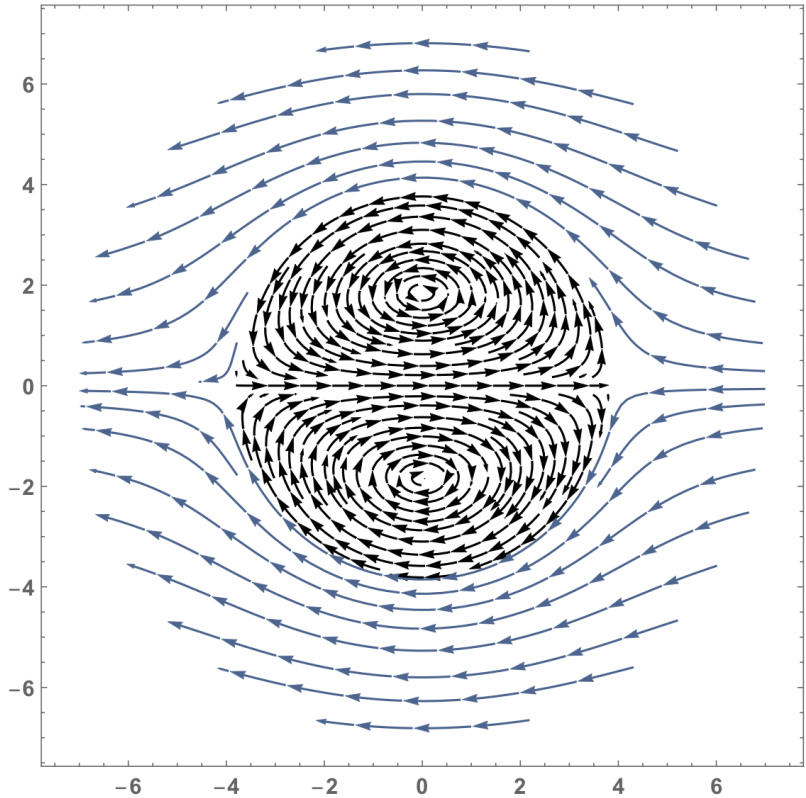

Figure 2. A streamline plot of $\psi_i$ and $\psi_e$ with the same parameters as Fig. 1.

For the Lamb-Chaplygin vortex dipole the first non-zero root of $J_1(b) = 0$ was used, but the boundary conditions will also be satisfied by the other higher-order roots of this equation which will allow higher-order vortices.[†] For example, $J_1(b) = 0$ is also satisfied for $b = 7.01559$ and $b = 10.1735$, the second and third zeros of the Bessel function $J_1$. Equation (9) then becomes

$$\psi_i = \frac{2U}{J_0(b_i)} J_1(r) \sin\theta \qquad a_{i-1} \leq r \leq a_i,$$

(11)

where $a_0 = 0$ and $a_i$, for $i = 1, 2, 3, \ldots$ has been set equal to $b_i$ so that the cylindrical boundary has a radius equal to $b_i$; as mentioned above there is no longer a need to distinguish the two. The parametric and streamline plots of $\psi_i$ for $i = 1, 2, 3$ are shown in Fig. 3.

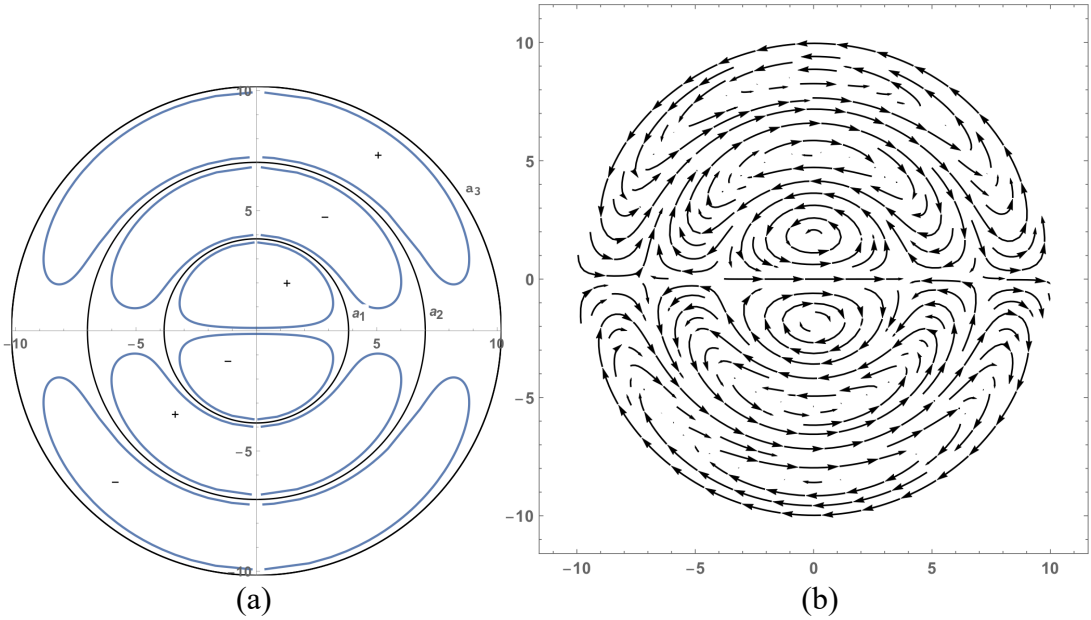

(a)          (b)

Figure 3. A parametric and stream plot for Eq. (9) for $i$ = 1, 2, 3. (a) The circles in the parametric plot designate the cylindrical "boundaries" that are the solutions of $J_1(r) = 0$; the Lamb-Chaplygin dipole vortex appears inside the circle at $a_1 = b_1$. The values of $\psi$ chosen for the plot are such that $|\psi| = 0.1$. Using $i = 2$ gives the Lamb-Chaplygin vortex and the first higher-order vortex between $a_1$ and $a_2$ without the cylindrical boundary at $a_1$ and similarly for $i = 3$. The streamline plot in (b) for $0 \leq \theta \leq 2\pi$ shows the handedness of the vorticity with increasing $r$. These are indicated in (a) by + and – signs. The divisions between the stream plots as in (a) occur at $J_1(r) = 0$.

Since only $J_1$ is used in Eq. (11), the solutions shown in Fig. 3 could be matched by the boundary conditions given above to a uniform flow for $r \geq b_i$.

---

[†] The designation of "higher-order" with respect to vortices comes from the relation to Bessel function order.

## The asymmetric Chaplygin vortex

By taking the relation between the vorticity and stream function to be $\omega = \frac{b^2}{a^2}(\psi - \lambda)$, Chaplygin found another solution to Eq. (1):

$$\psi_i = \frac{2Ua}{b\,J_0(b)} J_1\left(\frac{br}{a}\right) \sin\theta - \lambda\left[1 - \frac{J_0\left(\frac{br}{a}\right)}{J_0(b)}\right] \qquad 0 \leq r \leq a,$$

(12)

where, to be consistent with the literature, the distinction between $a$ and $b$ has been reintroduced. Figure 4 shows a contour and streamline plot for the stream function given by Eq. 12. The boundary conditions are satisfied for any value of $\lambda$.

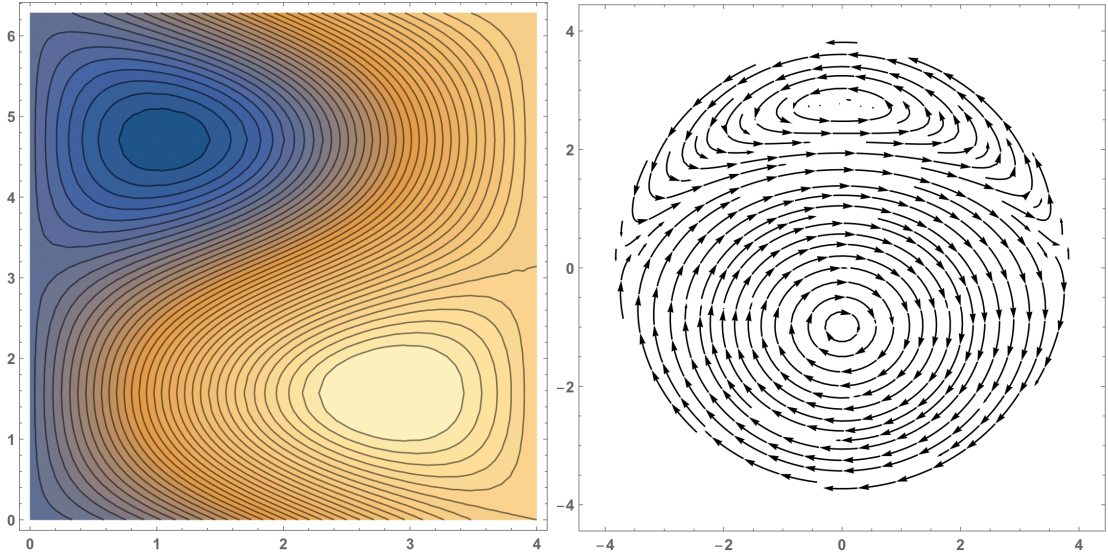

Figure 4. Contour and streamline plot for the asymmetric Chaplygin vortex given by Eq. 12 where $b$ corresponds to the first non-zero root of $J_1(b) = 0$. The value of $\lambda$ has been set equal to 0.3. Changing its sign reflects the figure about the zero line.

Using the linear relationship between the vorticity $\omega$ and the stream function $\psi$, the vorticity can be written as

$$\omega = \frac{2Ub}{a\,J_0(b)}\left[J_1\left(\frac{br}{a}\right)\sin\theta - \lambda\frac{b}{2Ua} J_0\left(\frac{br}{a}\right).\right]$$

(13)

For $r = a$, the vorticity becomes $\omega = -\lambda \frac{b^2}{a^2}$. Since the exterior vorticity for $r = a$ vanishes, there is a discontinuity in the vorticity across the circle $r = a$. Henceforth, $a$ and $b$ will be reidentified. For the first non-zero root of $J_1(b) = 0$, a contour plot of the vorticity is given in Fig. 5; Figure 6 shows a higher-order contour and streamline plot for the second non-zero root of $J_1(b) = 0$.

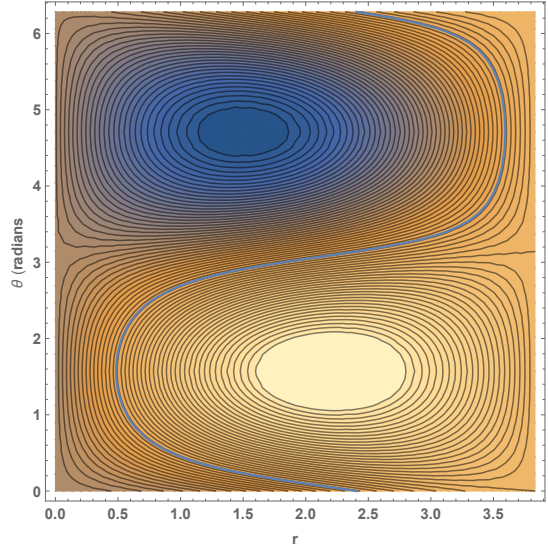

Figure 5. A contour plot for the vorticity given by Eq. (13) for the first non-zero root of $J_1(b) = 0$. Here, the value of the parameter $\lambda$ for the contour plot is 0.1. The reverse S-shaped curve correspond to $\omega = 0$. The vorticity is positive for the lower plot $0 \leq \theta \leq \pi$ and negative for the upper. Because the value of $\lambda$ for the contour plot is smaller than that in Fig. 4 the asymmetry is not as great.

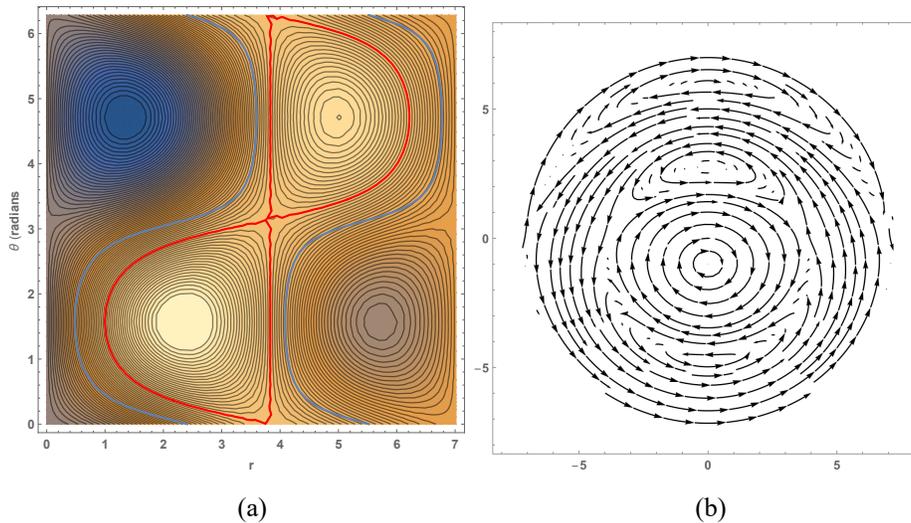

(a)          (b)

Figure 6. A contour plot for the stream function $\psi$ and a stream plot for the vorticity $\omega$ of the asymmetric Chaplygin vortex. These plots use $b$ corresponding to the second non-zero root of $J_1(b) = 0$, which is found at $r = 7.0155$. The value of $\lambda$ for the plots is 0.15. The vertical separation between the two dipole pairs in the contour plot occurs at $r = 3.8317$, the first non-zero root of $J_1(b) = 0$. The reverse S-shaped curves correspond to $\omega = 0$ and the red curve corresponds to $\psi = 0$.

Note the location of the vortices in the stream plot of Fig. 6 and the displacement of the central vortex from the zero-line due to the non-zero value of the asymmetry parameter $\lambda$.

**Generalization of the Lamb-Chaplygin and the asymmetric Chaplygin solution to the two-dimensional Euler equation for inviscid incompressible flow**

The derivation of the solution given in Eq. (12) is non-trivial and does not seem to appear in the available literature. I have directly confirmed that it is indeed an actual solution. I have also found a generalization given by

$$\psi = J_n(r)\,Sin(n\theta) + \lambda\big(1 - J_0(r)\big).$$
(14)

There is also a known generalization of the Lamb-Chaplygin solution that was used to produce the two more complex figures in the Introduction given by

$$\psi = J_n(r)\,Sin(n\theta).$$
(15)

These solutions look relatively simple since they do not include any constants need to match any imposed boundary conditions. A streamline plot for Eq. (15) for $n = 3$ and $\lambda = 0.3$ is shown below.

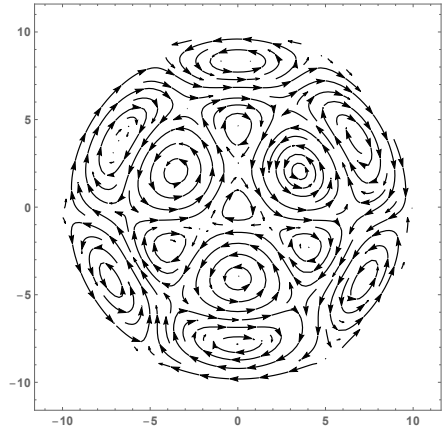

Linear combinations of such solutions are also solutions because of a fundamental property of second order inhomogeneous linear partial differential equations: If $L$ is a linear differential operator, and $\psi_1, \psi_2, \ldots \psi_N$ are solutions of the inhomogeneous equation $L\psi_j = f_j$, where $f_i$ is some function. Then a linear combination $\Psi$ of the $\psi_j$ is a solution of $L\Psi = \sum_{j=1}^{N} c_j\,f_j$. Not surprisingly, this is known as the principle of superposition for second order partial differential equations.

Could these solutions be useful for understanding atmospheric phenomena such as blocking? This might be the case if boundary conditions do not restrict the Bessel function to $n = 1$. Those used above restricting the Bessel function to $J_1$ came from the flow around what could be a physical boundary cylinder in order to obtain a steady flow. For such a boundary the normal component of the velocity of the streamlines vanishes when the surface is stationary. The relation between physical bounding surfaces and streamlines or stream surfaces comes from the fact that for a steady flow a streamline or surface may be replaced by a physical boundary, without affecting the flow, since the normal component of the velocity at a stream surface vanishes when the surface is stationary.

When the boundary is not stationary, the normal component of the fluid velocity at the boundary must equal the component of the boundary velocity along its own normal. That is, the boundary condition amounts to $DF/Dt = 0$, where the operator D represents differentiation with respect to the motion (or following the motion) and $F$ is the equation of the boundary surface $F(x,y,z,t) = 0$.

When such dynamic boundary conditions result from two different fluids, pressure intensity (difference in pressure) must change continuously across the boundary. In the case of interest here, the pressure intensity is that associated with an atmospheric vortex as the distance from its center increases so as to merge with surrounding streamlines; see Fig. 8 and the hydrodynamic analogous case shown by the Gulf Stream in Fig. 16.

**Atmospheric blocking and Lamb-Chaplygin vortices**
Atmospheric "blocking" occurs when the prevailing eastward winds (winds from the East known as westerlies) are blocked. In particular, it is the formation of a quasi-stationary vortex or vortex pair (often known as a modon) sufficiently strong to block the midlatitude storm track. Blocks caused by vortex dipoles are also known as diffluent blocks to distinguish them from omega blocks, so-named because they resemble an Ω. Omega blocks are not as persistent as dipole blocks. Blocks are also said to be due to "anomalous, persistent meandering of the jet stream".[6] When blocks appear, often at the edge of a continent, the jet stream splits trapping a vortex structure comprising the block.

A drawing of a dipole atmospheric block near Great Britain has been given by Voosen[7] in his News article in *Science* discussing the recent work of Wang and Kuang.[8] The figure is shown in Fig. 7 along with a differently configured diffluent block—note the asymmetry between the high- and low-pressure vortices in the second example.

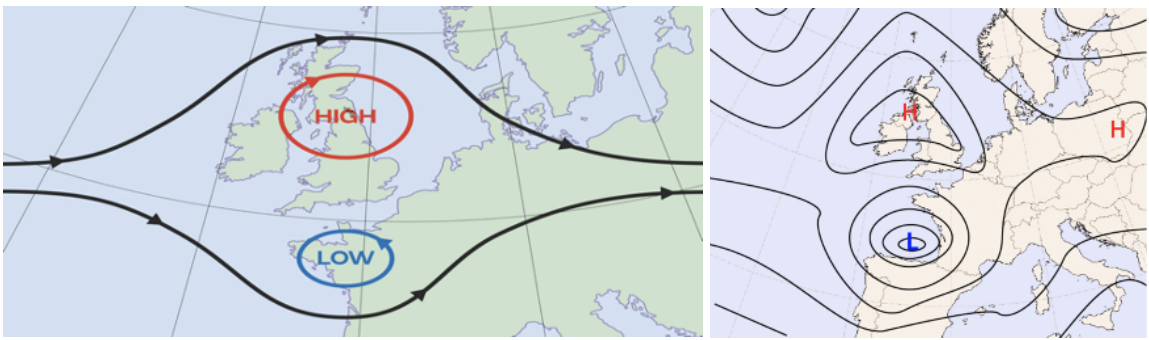

Figure 7.  Two dipole or modon atmospheric blocks. [Left: Diffluent block-metoffice.gov.uk; right: www.eumetrain.orgwww.eumetrain.org>satmanu>CMs>ScSh>backgr]

A satellite image of a diffluent block is shown in Fig. 8.

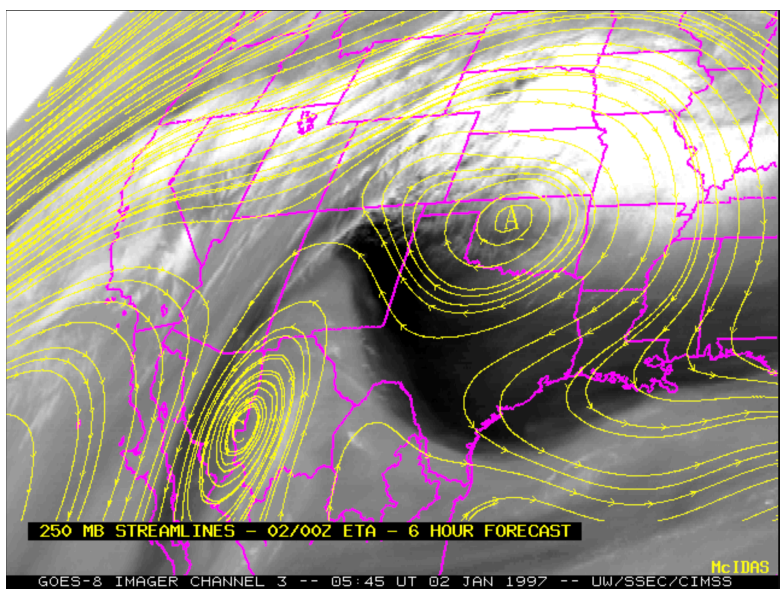

Figure 8.  A satellite image of a dipole block in the western US.
[Inside blocking boundary-https://cimss.ssec.wisc.edu/goes/misc/wv/ibb.html]

Note that the similarity of the symmetric blocks of Figs. 7 and 8 with Fig. 1. That Lamb-Chaplygin vortices could be important for understanding the phenomenon of atmospheric blocking was already raised by Haines and Marshall[9] in 1987. The essence of their Fig. 2 is essentially the same as Fig. 1 above.

Wang and Kuang used a two-layer quasi-geostrophic model on a $\beta$-plane to simulate the formation of a block. A discussion of the model and a weakly nonlinear model is given in their Appendices A and B. An example of their results is shown in Fig. 9.

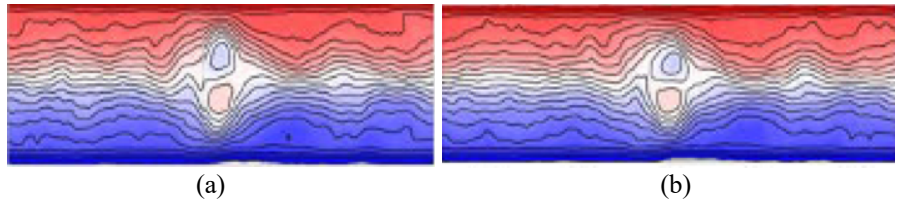

(a)                  (b)

Figure 9. Contour plot of the upper layer potential vorticity given by the quasi-geostrophic model. It is not clear whether the asymmetry of the vortices is real or an artifact of the simulation. [Adapted from Wang and Kuang's Fig. 2: arXiv: 1907.00999 (2019).

The highly asymmetric plot in Fig. 7 should be compared to the stream plot for Eq. 12 in Fig. 4 for the first non-zero root of $J_1(b) = 0$. Although distorted, the two are quite similar and suggest that higher order Chaplygin vortices could play a role in atmospheric blocks.

That the asymmetry is important in blocking has been confirmed by Luo,[10] but first a definition: a cyclonic shear is a horizontal wind shear such that it contributes to the cyclonic vorticity of the flow. In the Northern Hemisphere a cyclonic shear means that when facing downwind the wind speed increases from left to right across the flow. Luo developed what he called "a forced envelope Rossby soliton model in an equivalent barotropic beta-plane".[11] The model is based on two assumptions that make the synoptic-scale evolution equation linear while retaining the nonlinearity of the planetary-scale evolution equation.

Luo found that a resonant interaction between preexisting planetary and synoptic-scale waves can produce a dipole block but that its asymmetry, persistence, and intensity depend upon a cyclonic shear prior to the formation of the block. An example of such an antisymmetric block is shown in Fig. 10.

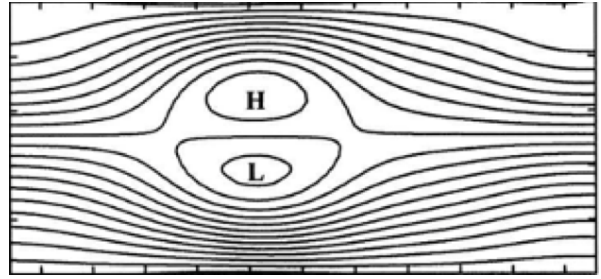

Figure 10. An antisymmetric block produced by the envelope Rossby soliton model. [Adapted from Luo *J. Atmos. Sci.*, **62** (2005), Fig. 3.]

**Higher-order Chaplygin vortices and blocking**

Higher-order Chaplygin vortices could play a role in the maintenance of atmospheric blocks by what is known as the selective absorption mechanism of Yamazaki and Itoh.[12] As a model for the synoptic eddies (small vortices) propagating in the waveguide formed by a jet stream consider the flow around a cylindrical obstruction. As shown in Fig. 11, small vortices develop behind the laminar separation points on each side of the obstruction. When the vortex reaches a threshold, it breaks away from the obstruction and moves downstream and another vortex of opposite vorticity forms behind the second separation point. This kind of wake, known as a von Kármán vortex trail, consists of equally spaced vortices of alternating vorticity. They propagate at a velocity, which is inversely proportional to the distance separating the vortices, that is somewhat less than the stream velocity.

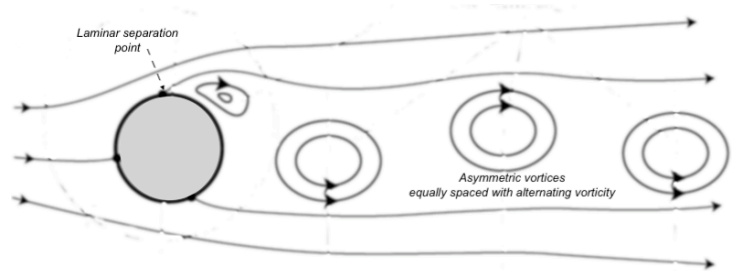

Figure 11. Flow around an obstructing cylinder for Reynolds number above ~100.

For Reynolds numbers below about 200, the stream of vortices can persist for large distances downstream.

Yamazaki and Itoh propose that vortex-vortex interactions are characterized by an attraction between anticyclonic vortices and a repulsion between cyclonic and anticyclonic vortices. In addition, that the vorticity of a large blocking anticyclone will extend to encompass a smaller anticyclone leading to their merger. There is a very clear analogy here with electromagnetism where the magnetic fields of two parallel wires carrying a current in the same direction attract each other.

Yamazaki and Itoh maintain that the alternating of the vorticity of the synoptic eddies is important for the maintenance, in particular, of a blocking dipole. This selective attraction mechanism allows a blocking dipole to attract synoptic eddies as long as these eddies are confined to the waveguide.[13] The series of vortices of alternating vorticity are then selectively absorbed by a blocking vortex thereby extending the lifetime of the block.

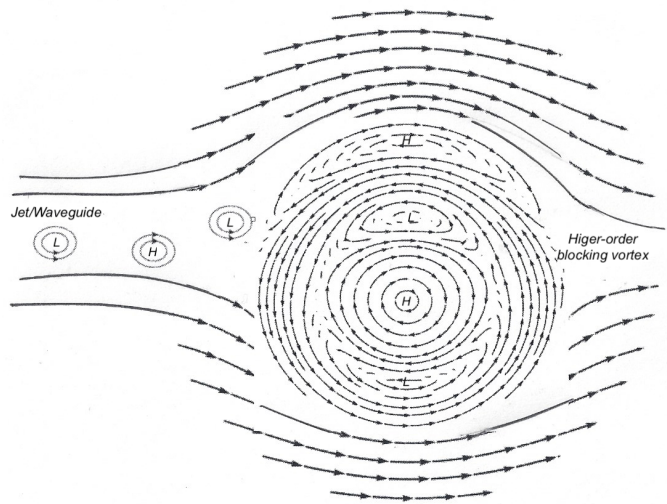

Figure 12. Synoptic eddies (the small vortices) propagating along jet/waveguide to be selectively absorbed by a higher-order antisymmetric blocking vortex that satisfies Eq. (12). This figure is for the Northern Hemisphere where the anticyclone has a clockwise flow of circulation.

In Fig. 12, a higher order vortex replaces the dipole block of Yamazaki and Itoh. The small low-pressure vortex (cyclone) entering the influence of the higher-order blocking vortex would merge with the low-pressure component of the blocking vortex; the next high-pressure vortex (anticyclone) in the jet/waveguide would merge with the top most high-pressure component of the blocking vortex.

The concept of selective absorption to extend the lifetime of atmospheric blocks was tested by Yamazaki and Itoh with case studies of some ten actual events where they confirmed that synoptic anticyclones are absorbed into blocking anticyclones, while synoptic cyclones are repelled by blocking anticyclones, and synoptic cyclones are repelled by blocking anticyclones and attracted to cyclonic blocks.

The study of interactions between vortices dates back to at least to the 1950 paper of Berson.[14] An adapted version of a figure from his study is shown in Fig. 13 to illustrate the actual complexity of vortices and vortex structure, and the difficulty of their interpretation. Even identifying vortices from atmospheric data in the presence of atmospheric shear is an ongoing subject of active research.[15]

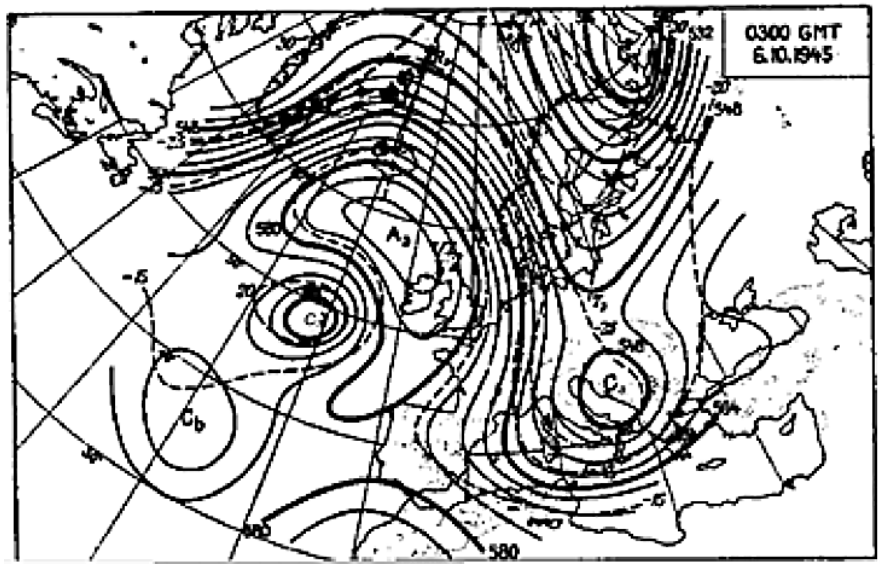

Figure 13. Contours and isotherms of the 500mb surface. Some illegible labels have been corrected. [Adapted from F. A. Berson, Fig. 3(d)]

In Fig. 13 the anticyclonic cell $A_3$ and the cyclonic cell $C_3$ constitute an anticyclonically rotating vortex pair. The interpretation of the sequence of such contours in Berson's Fig. 3 is that they show an evolution that leads to the growth of the cyclonic vortices $C_a$ and $C_b$.

The higher-order symmetric or antisymmetric blocking vortices that satisfies Eq. (12) are mathematical entities that, unlike the antisymmetric dipoles seen in Figs. 7 and 10, are unlikely to be directly observable in the atmosphere. The individual vortices making up the example in Fig. 6 would almost certainly rapidly be separated by the presence of atmospheric shear.

One approach that might help finding examples of real-world complex systems of blocking vortices comes from research conducted more than sixty years ago when it was realized that there is an analogy between the jet stream and ocean currents such as the Gulf Stream. The analogy was based on the correspondence between the horizontal potential temperature gradient of the Gulf Stream and the wind speed in the jet stream.[16] One example comes from the Hydrodynamics Laboratory of the University of Chicago where it was shown by Fultz[17] who demonstrated that narrow zonal currents could be formed in a rotating "dish pan" by heating the outer rim. The "Westerlies" near the rim correspond to atmospheric speeds of 50-150 mph. An adaptation of the relevant photograph is shown in Fig. 14.

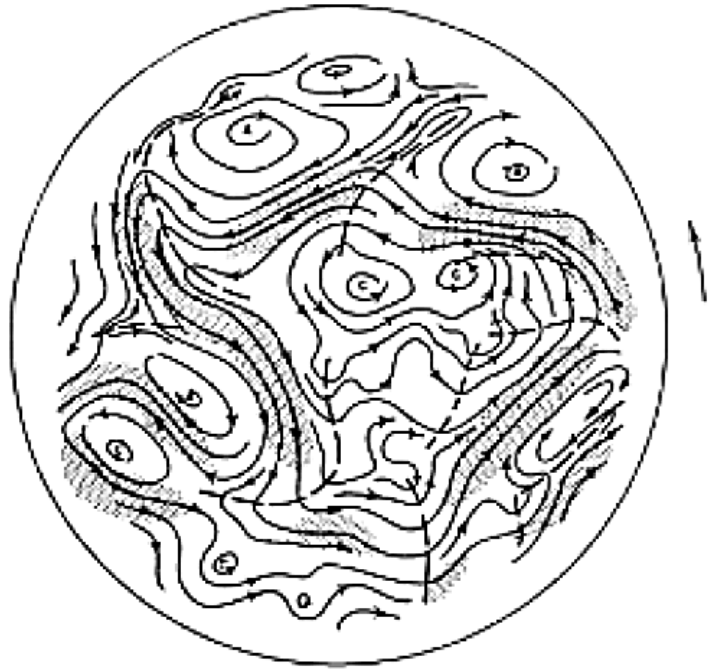

Figure 14. Streamlines of the relative circulation showing a well-defined wave pattern. Note the cyclonic and anticyclonic vortices designated by the labels C and A. Regions of high relative velocities are indicated by the shaded areas. The maximum speed is equivalent to a "westerly" wind of about 80 meters per second. (From records of the Hydrodynamics Laboratory, University of Chicago.)

If the shaded streamlines are identified with the jet stream, the C and A vortices in the lower left of the figure would correspond to a blocking dipole.

A more modern example showing that the Gulf Stream could be analogous to the jet stream is shown in Fig. 15. There the vortices labeled 1, 2, 3, and 4 constitute a complex blocking structure. As the pattern evolves, vortex 5 interacts with vortex 4 so as to increase the size of vortex 4. The evolution of these vortices is available in a video. Here is the link: https://svs.gsfc.nasa.gov/3913.

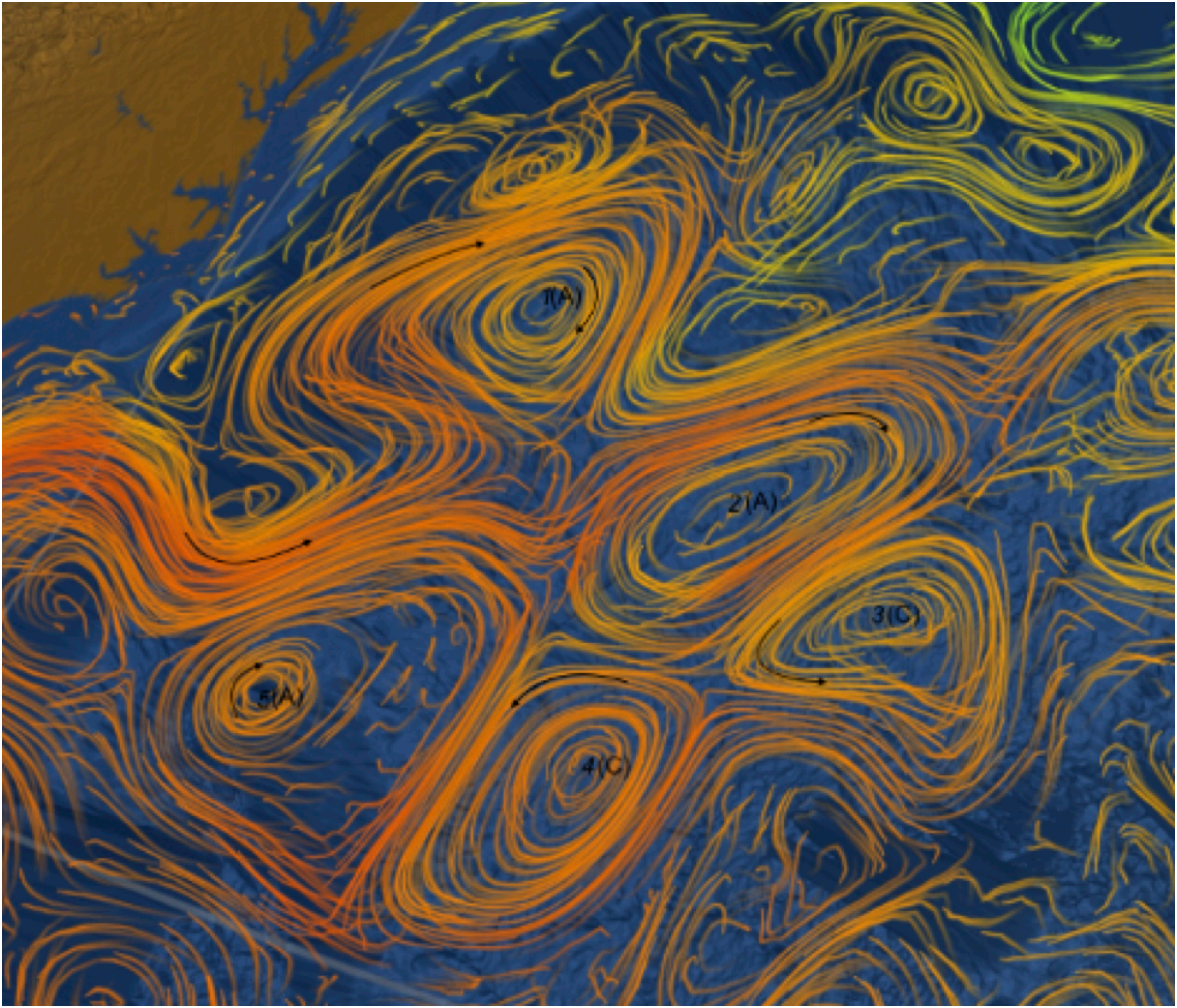

Figure 15. A portion of Gulf Stream sea surface currents and temperatures. The image is from an MIT/JPL project entitled Estimating the Circulation and Climate of the Ocean, Phase II (ECCO2). Vortices in the abstracted portion of the image have been numbered along with a C or A indicating cyclonic or anticyclonic rotation. The stream directions are indicated by the arrows. The Gulf Stream splits so as to encompass vortices 1-4. [NASA/Goddard Space Flight Center Scientific Visualization Studio]

While the vortices labeled 1, 2, 3, and 4 constitute a complex blocking structure, and have two cyclonic and two anticyclonic vortices such as the plot of the solution of Eq. (12) shown in Fig. (12), the evolution of these four vortices in Fig.15 does not clearly indicate a relation to those in Fig. (12). While vortices 2 and 4, or 2 and 3, could be considered as pairs making up a dipole, the past and future motion of the remaining two vortices does not bring them into a configuration with either pair that is fully consistent with Fig. (12). Nonetheless, the role of higher order Chaplygin vortices may be of value in the study of complex blocking structures.

More suggestive plots come from combining solutions. Figure 16 shows the combination of the solution that was used for Fig. 1 with an asymmetric solution.

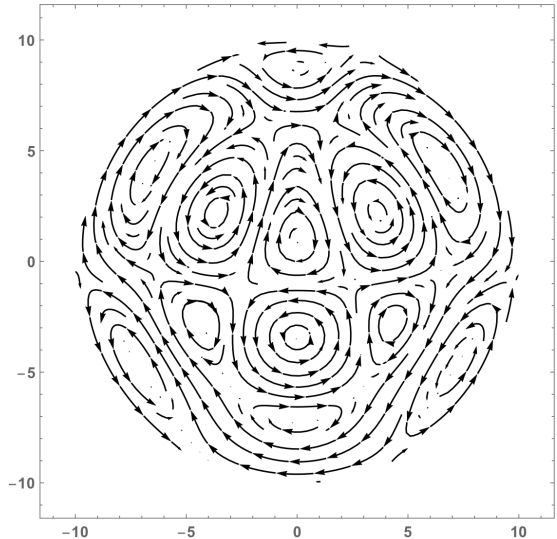

Figure 16. Using the principle of superposition for second order partial differential equations to combine the solutions given by Eqs. 11 and 12. Here $n = 3$ and $\lambda = 0.5$.

Note the strong relation between the central vortex dipole and that of the second vortex dipole of Fig. 7.

**Summary**

It has been shown that the equations describing both Lamb-Chaplygin and antisymmetric Chaplygin vortices have higher order solutions for vortices that could be of use in studying atmospheric blocking provided that the boundary conditions appropriate for a flow around a material cylinder are replaced with the non-stationary boundary conditions $DF/Dt = 0$, where the operator $D$ represents differentiation following the motion and $F$ is the equation of the boundary surface $F(x,y,z,t) = 0$. When this is the case, as the pressure intensity of an atmospheric vortex changes with increasing distance from its center its streamlines merge with those of the surrounding flow, as clearly shown in Figs. 14 and 15.

The character of the higher-order vortex structures was explored with plots of the streamlines as a function of the polar coordinates $r$ and $\theta$, parametric plots where the parameters are the numeric values of the stream function and the radius, and contour plots. The analogy between the jet stream and ocean currents such as the Gulf Stream was used to illustrate the real-world occurrence of both Lamb-Chaplygin vortices and complex antisymmetric Chaplygin vortices where vortex interactions become especially important.